# Coherent responses of resonance atom layer to short optical pulse excitation

Sergei O. Elyutin[*]

*Department of Physics, Moscow Engineering Physics Institute, Moscow 115409, Russia*

Coherent responses of resonance atom layer to short optical pulse excitation are numerically considered. The inhomogeneous broadening of one-photon transition, the local field effect, and the substrate dispersion are involved into analysis. For a certain intensity of incident pulses a strong coherent interaction in the form of sharp spikes of superradiation is observed in transmitted radiation. The Lorentz field correction and the substrate dispersion weaken the effect, providing additional spectral shifts. Specific features of photon echo in the form of multiple responses to a double or triple pulse excitation is discussed.

## 1. INTRODUCTION

A film with a thickness of the order of a wavelength placed on an interface of two dielectric media represents a simple example of 2D system, which optical properties has been intensively studied[1-6]. A thin film physics is attributed by an account of dipole-dipole interaction[7,8] in microscopic field acting on resonance atoms. The remarkable result in this field was the prediction of optical bistability[1,9-13] and the feasibility of the chaotic pulsation of the transmitted radiation[12]. The most of studies of the short pulse refraction by such nonlinear interface were carried out under the assumption that the atoms of a film were modeled by a two-level system in one-photon resonance with the incident radiation. Thus, a specific spatial synchronism of photon echo generated by a thin resonance layer on an interface was considered in[14]. The approaches worked out for conventional models were extended to another low dimension system, such as the layer of quantum dots[15], film of three level atoms[16,17] or thin film resonators filled with two-level medium[18], and another type of resonance, for instance, the two-photon resonance[19,20]

The purpose of this work is to demonstrate by numerical modeling, how a thin layer attributes, such as local field effect, a very small thickness of the film, the inhomogeneous broadening of resonance transition and even the dispersion of substrate material (Section 1) can affect the transmittance of layer and the temporal shape of transmitted ultra-short pulses (Section 2), including the observation of the photon echo effect [6,21,22] (Section 3). It turned out that Lorentz field correction plays important role in film structures, causing spectral dynamical shifts of atomic system observable in strong coherent coupling of pulsed radiation with a film.

## 1. A THIN FILM OF ATOMS ON AN INTERFACE

Let a thin film of atoms interacting resonantly with the electromagnetic field of light wave lie on the layer along the interface of two dielectric media in the $X$=0 plane. The dielectric media sur-rounding the film are characterized by the dielectric permittivity $\varepsilon_a$ at $x < 0$ and $\varepsilon_b$ at $x > 0$. Axis $Z$ is chosen in the plane of the interface. The resonance atoms are described by two-level model. The duration of the optical pulse is short in compare with polarization and population difference relaxation times, but at the same time, it is much longer than optical period. Therefore, the approximation of slowly varying complex amplitudes can be still applied. It is supposed that the thickness of the film $l$ is of the order or less than the wavelength of resonant radiation $l \sim \lambda$.

---
[*]E-mail: elyutin@mail.ru

Let the short pulse of the TE-wave be incident on the interface from the $x < 0$ side. The electromagnetic field vector $\boldsymbol{E}$ of such wave is orthogonal to the ZX plane for all $x$, i.e. the polarization vectors of the incident, refracted, and reflected wave are collinear to the Y axis:

$$\boldsymbol{E} = (0, E_y, 0), \quad \boldsymbol{H} = (H_x, 0, H_z).$$

The reflected wave returns to the $x < 0$ half-space but the refracted wave propagates into the $x > 0$ half-space. It is convenient to represent the field strength $\boldsymbol{E}$, and the polarization $\boldsymbol{P}$ of the resonant atoms inside the film as

$$\boldsymbol{E}(x,z,t) = \int\limits_{-\infty}^{\infty} \frac{d\omega}{2\pi} \frac{d\beta}{2\pi} \exp(-i\omega t + i\beta z) \tilde{\boldsymbol{E}}(x, \beta, \omega),$$

$$\boldsymbol{H}(x,z,t) = \int\limits_{-\infty}^{\infty} \frac{d\omega}{2\pi} \frac{d\beta}{2\pi} \exp(-i\omega t + i\beta z) \tilde{\boldsymbol{H}}(x, \beta, \omega),$$

$$\boldsymbol{P}(z,t) = \int\limits_{-\infty}^{\infty} \frac{d\omega}{2\pi} \frac{d\beta}{2\pi} \exp(-i\omega t + i\beta z) \tilde{\boldsymbol{P}}(\beta, \omega).$$

Outside the film the Fourier components $\tilde{E}(x, \beta, \omega)$ and $\tilde{H}(x, \beta, \omega)$ of the field vectors are determined by Maxwell's equations, and the components at $x = 0$ are determined by continuity condition so that for the TE case we obtain the system of equations

$$\frac{d^2 \tilde{E}}{dx^2} + \left(k^2 \varepsilon_j - \beta^2\right)\tilde{E} = 0, \quad \tilde{H}_x = -(\beta/k)E, \quad \tilde{H}_z = -\frac{i}{k}\frac{d\tilde{E}}{dx}, \quad \tilde{E}_y = \tilde{E} \tag{1}$$

with the boundary conditions

$$\tilde{E}(x = 0-) = \tilde{E}(x = 0+), \quad \tilde{H}_z(x = 0+) - \tilde{H}_z(x = 0-) = 4\pi i k \tilde{P}_y(\beta, \omega)$$

Here $j = 1, 2$, $k = \omega/c$ and $\beta$ is the propagation constant along the interface. Outside a thin film the solution of Eq. (1), with the allowance for the behavior of the field far from the film, has the form

$$\tilde{E}(x, \beta, \omega) = \begin{cases} \tilde{E}_{in} \exp(iq_1 x) + \tilde{E}_{ref} \exp(-iq_1 x), & x < 0 \\ \tilde{E}_{tr} \exp(iq_2 x), & x > 0 \end{cases}, \tag{2}$$

where $q_j = \sqrt{k^2 \varepsilon_j - \beta^2}$, $j = a, b$.

The boundary conditions at $x = 0$ provide the relations between the amplitudes of the incident $\tilde{E}_{in}$, reflected $\tilde{E}_{ref}$, and refracted (transmitted) $\tilde{E}_{tr}$ waves and the surface polarization $\tilde{P}_S = \tilde{P}_{\tilde{y}}$ of the film.

$$\tilde{E}_{tr}(\beta,\omega) = \frac{2q_a}{q_a + q_b}\tilde{E}_{in}(\beta,\omega) + i\frac{4\pi k^2}{q_a + q_b}\tilde{P}_S(\beta,\omega)$$

$$\tilde{E}_{r}(\beta,\omega) = \frac{q_a - q_b}{q_a + q_b}\tilde{E}_{in}(\beta,\omega) + i\frac{4\pi k^2}{q_a + q_b}\tilde{P}_S(\beta,\omega) \qquad (3)$$

It is convenient to introduce the notations for Fresnel transmission coefficient $T_{TE}$ and the coupling constant $\kappa_{TE}$

$$T_{TE}(\beta,\omega) = \frac{2q_1}{q_1 + q_2}, \;\; \kappa_{TE}(\beta,\omega) = \frac{4\pi k^2}{q_1 + q_2}.$$

For the given polarization of the film, relations (3) determine the field in entire space. It should be emphasized that (3) in no way are related with the assumption of slowly varying envelopes of the optical pulses and are exact.

We shall now focus our attention on the refracted wave. Let consider only the case $\varepsilon_a < \varepsilon_b$, where the total internal reflection does not occur for any angle of incidence $\varphi = \arccos(q_a / k\sqrt{\varepsilon_a})$, and Snell's law sets the refraction angle $\psi$ :

$$\sin\psi = (\beta / k\sqrt{\varepsilon_b}) = \sqrt{\varepsilon_a / \varepsilon_b}\sin\varphi. \qquad (4)$$

We concentrate on the case of TE waves. The Fourier components of the amplitudes of the macroscopic fields are coupled by the equations (3), providing an exact relationship:

$$\tilde{E}_{tr}(\beta,\omega) = T(\beta,\omega)\tilde{E}_{in}(\beta,\omega) + i\kappa(\beta,\omega)\tilde{P}_S(\beta,\omega). \qquad (5)$$

Indices TE are omitted for brevity.

The surface polarization $\tilde{P}_S(\beta,\omega)$ is determined by the thin layer atom response. It is generated by the local field $\tilde{E}_L(\beta,\omega)$, which, in its turn, is the sum of the field in the film $\tilde{E}_{tr}(\beta,\omega)$ and the bulk polarization $\tilde{P}(\beta,\omega)$ [1,11]

$$\tilde{E}_L(\beta,\omega) = \tilde{E}_{tr}(\beta,\omega) + 4\pi\alpha\tilde{P}(\beta,\omega). \qquad (6)$$

Parameter $\alpha$ accounts the effect of environment. It is often chosen $\alpha = 1/3$ for isotropic distribution of atoms in bulk material.

An optical pulse incident on the film presents in the form of a quasi-harmonic wave

$$E(x,z,t) = \mathcal{E}(x,z,t)\exp[-i\omega_0 t + i\beta_0 z],$$

where $\mathcal{E}(x,z,t)$ is the pulse envelope varying slowly in space and in time, $\beta_0 = \beta(\omega = \omega_0)$ is the propagation constant at the frequency of the pumping wave.

The condition of the quasi-harmonicity can be expressed in terms of Fourier components of electric field $\tilde{E}(\omega,\beta)$ the pulse and its shape $\mathcal{E}(\omega,\beta)$. This condition writes as

$$\tilde{E}(\omega_0 + \omega, \beta_0 + \beta) = \mathcal{E}(\omega,\beta)$$

with $\omega \ll \omega_0$ and $\beta \ll \beta_0$.

Thus, under the approximation of quasi-harmonic waves the coupling equations (3) re-write as

$$\mathcal{E}_{tr}(\beta,\omega) = T(\beta_0 + \beta, \omega_0 + \omega)\mathcal{E}_{in}(\beta,\omega) + i\kappa(\beta_0 + \beta, \omega_0 + \omega)\mathcal{P}_S(\beta,\omega), \qquad (7a)$$

$$\mathcal{E}_r(\beta,\omega) = R(\beta_0 + \beta, \omega_0 + \omega)\mathcal{E}_{in}(\beta,\omega) + i\kappa(\beta_0 + \beta, \omega_0 + \omega)\mathcal{P}_S(\beta,\omega). \qquad (7b)$$

In (7) the Fourier component of the slowly varying surface polarization $\mathcal{P}_S(\beta,\omega)$ is the functional of the Fourier component of the local field envelope $\mathcal{E}_L(\beta,\omega)$:

$$\mathcal{E}_L(\beta,\omega) = \mathcal{E}_{tr}(\beta,\omega) + 4\pi\alpha \, \mathcal{P}(\beta,\omega). \qquad (8)$$

Here $\mathcal{P}(\beta,\omega)$ is the Fourier component of the bulk polarization of the film material.

In order to take into account the finite spectral width of the wave packet, which formed a quasi-harmonic signal, one can expand the coefficients $T(\beta_0 + \beta, \omega_0 + \omega)$, $R(\beta_0 + \beta, \omega_0 + \omega)$ and $\kappa(\beta_0 + \beta, \omega_0 + \omega)$ in power series of $\omega$ until several first terms of these series.

For the sake of simplicity let us assume that the incident wave is isotropic in the plane of interface, so the terms proportional to $\beta/\beta_0$ in the expansion of $T$, $R$ and $\kappa$ are ignored.

Then one can write with the accuracy up to the second power of $\omega$:

$$T(\beta_0 + \beta, \omega_0 + \omega) \approx T(\beta_0,\omega_0) + T_1(\beta_0,\omega_0)\omega + T_2(\beta_0,\omega_0)\omega^2,$$

$$R(\beta_0 + \beta, \omega_0 + \omega) \approx R(\beta_0,\omega_0) + R_1(\beta_0,\omega_0)\omega + R_2(\beta_0,\omega_0)\omega^2.$$

Only the first term of a corresponding series is kept for the coupling constant, i.e. $\kappa(\beta_0 + \beta, \omega_0 + \omega) \approx \kappa(\beta_0,\omega_0)$.

With this account, the coupling equations (7) allow to transfer from the spectral presentation to the dynamic presentation:

$$\mathcal{E}_{tr}(z,t) = \left( T_0 + iT_1\frac{\partial}{\partial t} - T_2\frac{\partial^2}{\partial t^2} \right)\mathcal{E}_{in}(z,t) + i\kappa\mathcal{P}_S(z,t), \qquad (9a)$$

$$\mathcal{E}_r(z,t) = \left( R_0 + iR_1\frac{\partial}{\partial t} - R_2\frac{\partial^2}{\partial t^2} \right)\mathcal{E}_{in}(z,t) + i\kappa\mathcal{P}_S(z,t). \qquad (9b)$$

The notations introduced in (9) are:

$$T_0 = \frac{2\cos\varphi}{\cos\varphi + \sqrt{\cos^2\varphi + (\varepsilon_b - \varepsilon_a)/\varepsilon_a}} \qquad \kappa = \frac{4\pi(\omega/c)}{\cos\varphi + \sqrt{\cos^2\varphi + (\varepsilon_b - \varepsilon_a)/\varepsilon_a}} \qquad (10)$$

Let atoms or molecules, whose resonance levels are coupled by one photon transition at the frequency of pulse carrier, form a thin layer.

The slow varying matrix elements of density matrix satisfy Bloch equations

$$\frac{\partial \sigma_{\Delta\omega}}{\partial t} = i\Delta\omega\sigma_{\Delta\omega} + i\Omega n_{\Delta\omega}, \qquad \frac{\partial n_{\Delta\omega}}{\partial t} = \frac{i}{2}(\sigma_{\Delta\omega}\Omega^* - \sigma_{\Delta\omega}^*\Omega),$$

where $\Omega = d\mathcal{E}_L/\hbar$ is an instant Rabi frequency, $d$ is a matrix element of the operator of dipole moment of resonance transition. The variables $n_{\Delta\omega}$ and $\sigma_{\Delta\omega}$, depending on the detuning $\Delta\omega$, are related to the matrix elements of the density matrix by the relationships $n = \rho_{11} - \rho_{22}$, $\rho_{12} = \sigma\exp(-i\omega_0 t + i\beta_0 z)$.

The surface polarization $\mathcal{P}_S(z,t)$ expresses in terms of these quantities as

$$\mathcal{P}_S(z,t) = d\,n_a l < \sigma_{\Delta\omega}(z,t) >,$$

where $n_a$ is the bulk density of resonance atoms, the angular brackets means a summation over the frequency detuning $\Delta\omega$ of resonance frequency of individual atoms from the center of the inhomogeneous line.

Referring to equations (8), (9), formulae (10) and the relationship

$$\mathcal{P}(z,t) = d\,n_a < \sigma_{\Delta\omega}(z,t) >,$$

the instant Rabi frequency is written in the following form

$$\Omega = \frac{4\pi n_a d^2}{\hbar}\left(\alpha + i\frac{\kappa l}{4\pi}\right) < \sigma_{\Delta\omega} > + \left(T_0 + iT_1\frac{\partial}{\partial t} - T_2\frac{\partial^2}{\partial t^2}\right)\frac{d}{\hbar}\mathcal{E}_{in}(z,t). \qquad (10)$$

Following[1,2], the "co-operative time" $t_c = \hbar\left(4\pi n_a d^2\right)^{-1}$ can serve as the time scale of the problem. Then the main equations of the model write

$$\frac{\partial \sigma_x}{\partial \tau} = i\gamma x\sigma_x + ie_{loc}n_x, \quad \frac{\partial n_x}{\partial \tau} = \frac{i}{2}(\sigma_x e_{loc}^* - \sigma_x^* e_{loc}), \qquad (11a)$$

$$e_{loc} = f(\tau) + (\alpha + ig\chi) < \sigma_x >, \qquad (11b)$$

where $x = \Delta\omega T_2^*$, $\gamma = t_c/T_2^*$, $\chi(\varphi) = 2\pi\left(\sqrt{\varepsilon_a}\cos\varphi + \sqrt{\varepsilon_a\cos^2\varphi + (\varepsilon_b - \varepsilon_a)}\right)^{-1}$, $\tau = t/t_c$, thickness factor $g = l/\lambda$, $\lambda$ is the wavelength of the carrier wave, $e_{loc(in)} = d\mathcal{E}_{loc(in)}\hbar^{-1}t_c$, $T_2^*$ is the inhomogeneous dephasing time, $< .... >$ means averaging over the radiators $x$ within the inhomogeneous resonance absorption line, and the field penetrated the layer is

$$f(\tau) = \left( \mathcal{T}_0 + i\mathcal{T}_1 \frac{\partial}{\partial \tau} - \mathcal{T}_2 \frac{\partial^2}{\partial \tau^2} \right) e_{in}, \quad \mathcal{T}_0 = T_0, \quad \mathcal{T}_1 = T_1/t_c, \quad \mathcal{T}_2 = T_2/t_c^2.$$

The macroscopic field

$$e_{tr} = f(\tau) + ig\chi < \sigma_x > \tag{12}$$

is a detectable field spreading beyond the interface.

The characteristic for the two-level systems "co-operative" time $t_c = \hbar \left( 4\pi n_a d^2 \right)^{-1}$ is the time the polarization is induced by the field of the travelling pulse. If to choose the dipole moment $d \sim 10^{-18}$ CGSE and the density of resonance atoms $n_a \sim 10^{18}$ cm$^{-3}$, then $t_c \approx 8 \cdot 10^{-11}$s. The characteristic field strength can be determined from the condition $d\mathcal{E}_{char} \hbar^{-1} t_c \approx 1$. $\mathcal{E}_{char} = 4\pi n_a d = \hbar \left( dt_c \right)^{-1} \approx 10$ CGSE. For comparison, the intra-atomic field can be estimated as $\mathcal{E}_{atom} \approx 10^6$ CGSE. The corresponding peak intensity of the pulses, illuminating the film, is $I_{char} = c\mathcal{E}^2/8\pi \approx 10^4$ W·cm$^{-2}$. The Lorentz correction provides $\mathcal{E}_L = 4\pi n_a \alpha d = \alpha \mathcal{E}_{char} \approx 3$ CGSE. The strength of the field of radiation from the atom layer is $\mathcal{E}_{rad} = \left( 4\pi \right)^{-1} \kappa l \cdot \hbar \left( dt_c \right)^{-1} = \sqrt{\varepsilon_a} \chi g \mathcal{E}_{char} \approx 20 \div 30$ CGSE.

So the field equation (11b) re-writes in a straightforward form ( $\varepsilon_a = 1$ ):

$$e_{loc} = \left( \mathcal{T}_0 + i\mathcal{T}_1 \frac{\partial}{\partial \tau} - \mathcal{T}_2 \frac{\partial^2}{\partial \tau^2} \right) e_{in} + (e_L + ie_{rad}) < \sigma_x >, \tag{13}$$

where $e_{L(rad)} = \mathcal{E}_{L(rad),} / \mathcal{E}_{char}$ .

The dispersion coefficients $\mathcal{T}_1$ and $\mathcal{T}_2$ in the combined transmittivity can be estimated by the inequalities $\mathcal{T}_1/\tau_p < \mathcal{T}_0$, $\mathcal{T}_2/\tau_p^2 < \mathcal{T}_0$, where $\tau_p$ is the dimensionless duration of an excitation pulse. As the Fresnel coefficient $\mathcal{T}_0 \approx 1$, then for the pulses of several units in duration the values of $\mathcal{T}_1$ and $\mathcal{T}_2$ can be relatively large and be both positive and negative sign.

# 3. OPTICAL PULSE TRANSMISSION THROUGH A THIN FILM OF RESONANCE ATOMS

In the subsequent calculations the pulse incident the film are assumed having either the *sech* form: $e_m \sec h\left( (\tau - \tau_0) \tau_p^{-1} \right)$ or in the form of smooth step

$$e_m/2\Big(\tanh\big((\tau-\tau_0)\tau_f^{-1}\big)-\tanh\big((\tau-\tau_0-\tau_p)\tau_f^{-1}\big)\Big),$$

where $\tau_0$, $\tau_p$, and $\tau_f$ are the pulse delay, pulse duration, and pulse edge duration time respectively.

In this section, the resonance absorption line is supposed to be narrow and $\delta$ represents a certain detuning from exact resonance. Then equations (11) can be rewritten as

$$\frac{\partial\sigma}{\partial\tau}=i(\delta+\alpha n)\sigma+ine_{tr}, \ \frac{\partial n}{\partial\tau}=\frac{i}{2}(e_{tr}^*\sigma-\sigma^*e_{tr}), \ e_{tr}=f(\tau)+ig\chi\sigma, \ \gamma x=\delta. \quad (14)$$

## A. Pulse shape

To obtain simple formulae let the Lorentz field effect be absent $\alpha=0$. Then, equations of the model (11) at exact resonance ($\delta=0$) yields the solution

$$\sigma=i\sin\Theta, \ n=\cos\Theta, \ \Theta=\int_0^\tau e_{tr}(\tau')d\tau'$$

In terms of Bloch angle $\Theta$ the coupling equation (11b) or (13) transform to

$$\frac{\partial\Theta}{\partial\tau}+e_{rad}\sin\Theta=\mathcal{S}_0 e_{in}(\tau) \quad (15)$$

The solution of (15) was found in[4] for a rectangular pulse form. In the case $e_m\ll 1$ the field of the transmitted pulse looks like

$$e_{tr}=\begin{cases} -\mathcal{S}_0 e_m\big[\exp(e_{rad}\tau)-1\big]\exp(-e_{rad}\tau), & \tau>\tau_p \\ \mathcal{S}_0 e_m\exp(-e_{rad}\tau), & 0\le\tau\le\tau_p \,. \end{cases} \quad (16)$$

To illustrate this, the variant of numerical simulation of (11) for a smooth step input pulse is depicted in Fig. 1(a). We neglect the substrate dispersion at this stage. It is seen that the weak pulse is almost totally reflected.

If the amplitude of the exciting field is not too large, than the field is in resonance with the quantum system and so the pulse is strongly reflected from the layer as it was discussed in a number of papers[1,2,4]. The exponential tails at the leading and the trailing edge of the pulse are the signals of free induction decay.

The calculation run with the Lorentz factor $\alpha\approx 1$ revealed the role of the local field[7] as the additional detuning causing the shift from resonance and the loss of coherency. That leads to practically total transparency of the film for pulse radiation (Fig. 1(a)).

In the intermediate case, $\mathcal{S}_0 e_m\le 1$ the solution of equations (14) in the interval $0\le\tau\le\tau_p$ contains characteristic hyperbolic functions[4]:

$$e_{tr}(\tau) = \mathcal{T}_0 e_m Q_1^2 \left( Q_1 sh(e_{rad} Q_1 \tau) + ch(e_{rad} Q_1 \tau) - \mathcal{T}_0^2 e_m^2 \right)^{-1},$$

where $Q_1^2 = 1 - \mathcal{T}_0^2 e_m^2$.

The step pulse (Fig. 1(b)) splits into a series of spikes whose number, similar to the SIT[23] phenomenon, grows with the increase of $\Theta(\tau_p)$. The spikes on the profile of the transmitted signal are the pulses of superradiation arose due to the pendulum motion of the Bloch vector. The pendulum quickly passes the point of equilibrium while moves slowly near non-equilibrium position. In Fig. 1(b), the population of the ground level exhibits Rabi oscillation. That means that the Bloch vector makes the full rotation so that the population restores to the beginning of the next peak. We observe a SIT-like splitting of the powerful incident pulse into several sub-pulses of superradiation with a corresponding oscillation of population.

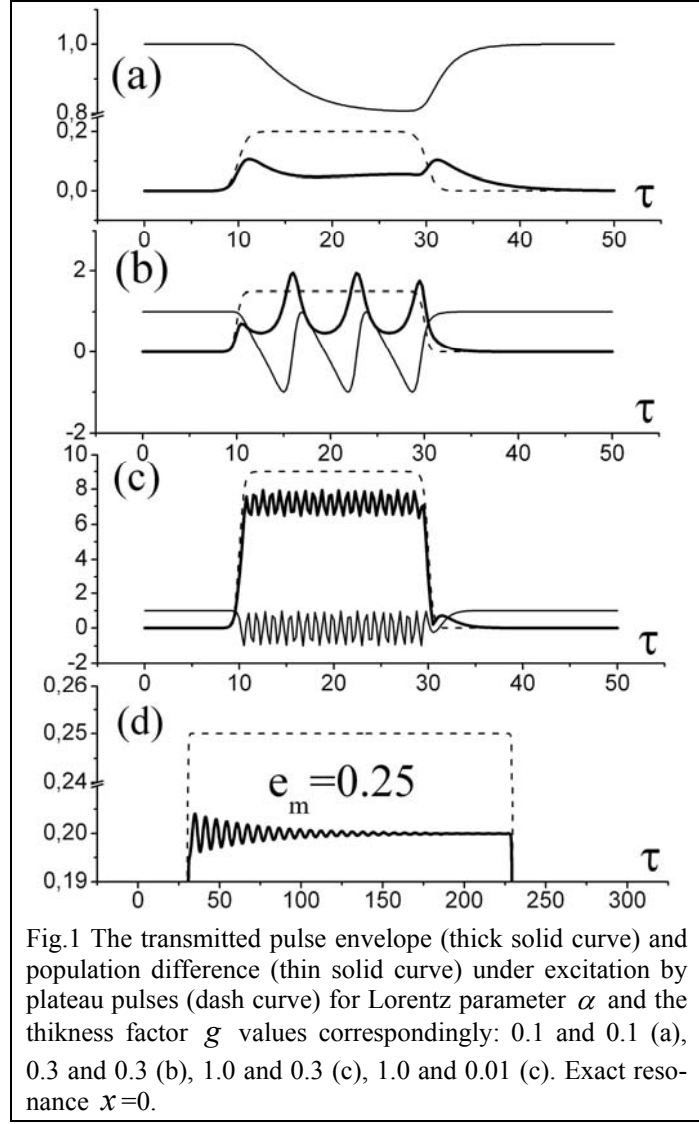

Fig.1 The transmitted pulse envelope (thick solid curve) and population difference (thin solid curve) under excitation by plateau pulses (dash curve) for Lorentz parameter $\alpha$ and the thikness factor $g$ values correspondingly: 0.1 and 0.1 (a), 0.3 and 0.3 (b), 1.0 and 0.3 (c), 1.0 and 0.01 (c). Exact resonance $x = 0$.

At last, for the strong fields, when $\mathcal{T}_0 e_m > 1$ at time interval $0 \le \tau \le \tau_p$ the corresponded solution of the reduced problem (14) has the oscillatory form (Fig.1(c)):

$$e_{tr}(\tau) = \mathcal{T}_0 e_m Q_2^2 \left( \mathcal{T}_0^2 e_m^2 + Q_2 \sin(e_{rad} Q_2 \tau) - \cos(e_{rad} Q_2 \tau) \right)^{-1},$$

where $Q_2^2 = \mathcal{T}_0^2 e_m^2 - 1$.

All three cases are common in that the dynamics of transmission of short pulses is determined by parameter $e_{rad}$, responsible for effectiveness of radiation interaction with the layer of resonance atoms.

It is interesting that in the case of weak interaction (Fig. 1(d)) the quasi-continuous excitation causes the response in the form of damping oscillations tending to a stationary value[16], but only when the reaction of the film is weak ($g < 1$).

**B. Transmission coefficient**

If to introduce the transmission coefficient as a square root of the ratio of the transmitted pulse energy to the incident pulse energy then the transmission coefficient for the weak rectangular pulses can be derived after some algebra from (16):

$$\mathcal{T}_{tr} = \mathcal{T}_0 \tau_p^{-1/2} \left(1 - \exp\left(-\tau_p\right)\right)^{1/2} \qquad (17)$$

Expression (17) gives $\mathcal{T}_{tr} = \mathcal{T}_0 \left(1 - \tau_p/2\right)^{1/2}$ for $\tau_p \ll 1$, and $\mathcal{T}_{tr} = \mathcal{T}_0 \tau_p^{-1/2}$ for $\tau_p \gg 1$. That means that a film of resonance atoms serves as a mirror for the long weak pulses.

In other cases the numerical analysis seemed to be more adequate. Each point on graphics (Fig. 2) is the result of solution of the whole system (11) with a subsequent integration of the solution over time. The pulse shapes for the final pulse duration for each of three incoming amplitude in Fig. 2(a,b,c) are depicted on the corresponded panels in Fig. 1.

In Fig. 2(a) the graphic is placed of the transmission coefficient $\mathcal{T}_{tr}$ vs incident

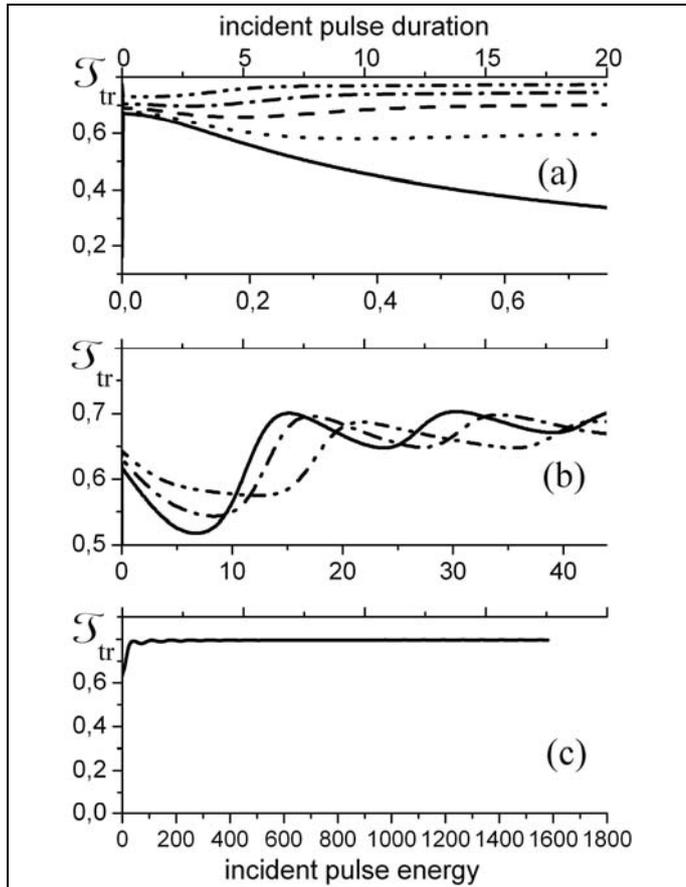

Fig.2 Transmission coefficient vs incident pulse energy for weak (a), medium (b) and strong (c) input fields. (a) factor $g$ =0.1, parameter $\alpha$ values: solid curve – 0.0, dot curve – 0.3, dash curve – 0.5, dash-dot curve – 0.7, dash-dot-dot curve – 1.0. (b) $g$ =0.3, solid – 0.0, dash-dot – 0.7, dash-dot-dot – 1.0. (c) $g$ =0.3, same curve for any $\alpha$ from the above set. Exact resonance $x$ =0.

pulse energy at the constant pulse amplitude $e_{in} = 0.2$. The solid curve exhibits the anticipated reflection of a weak pulse (17). The transient regime ($e_m$ =1.5) is featured by the oscillation of the transmittivity (Fig. 2(b)) due to the arising of new peaks on the temporal profile of a transmitted pulse (Fig. 1(b)). The transparency of the resonance atom layer reaches the Fresnel limit

$\mathcal{T}_0$ (Fig. 2(c)) and it does not practically alter with the energy growth when the amplitude of the input pulses is sufficiently large (Fig. 1(c)).

## C. Local field effect in coherent transmission

Lorentz field contribution does have a noticeable effect only for weak and intermediate pulse amplitudes when the action of the external field and the Lorentz field correction can be compared. Then the presence of the local field is accounted as an additional detuning, which drives the quantum system out from resonance. For a strong pulse excitation, the contribution of the local field effect is negligible (Fig. 1(c)).

The 3D plot in Fig. 3(a) presents the evolution of the pulse of medium height under the sweep of detuning $\delta$ from its negative to positive values. As it has been shown above, when the Lorentz field effect is weak ($\alpha < 1$), the step pulse suffers a strong coherent interaction with the layer accompanying by the formation of sharp spikes of superradiation (Fig. 3(b)) with a remarkable satellite spike in the later time moments.

The reason for the decline and

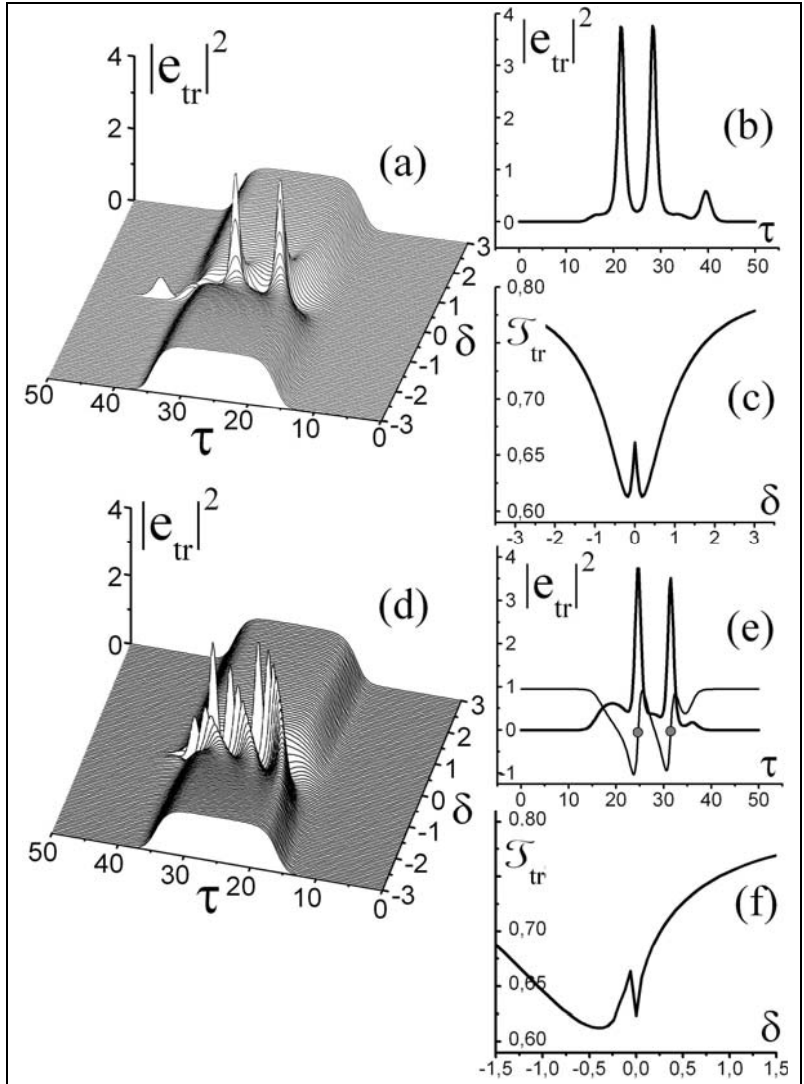

Fig.3 Evolution of pulse shape of medium height under the sweep of $\delta$. $\mathcal{T}_1 = 0$, $\mathcal{T}_2 = 0$. (a) $\alpha = 0$; (b) $\alpha = 0$, $\delta = 0$; (c) $\alpha = 0$; (d) $\alpha = 1$; (e) transmitted pulse envelope (thick solid curve), total detuning $\delta_{tot}$ population difference (thin solid curve), $\alpha = 1$, $\delta = 0$; (f) $\alpha = 1$.

the subsequent growth of $\mathcal{T}_{tr}$ vs $\delta$ in Fig.3(c) is the change of resonance conditions of field interaction with layer, while a sudden narrow maximum in the bottom of the cited plot is the result of the blooming of resonance film by intensive peaks of radiation on the contour of transmitted field (see Figs. 1(c), 2(c)).

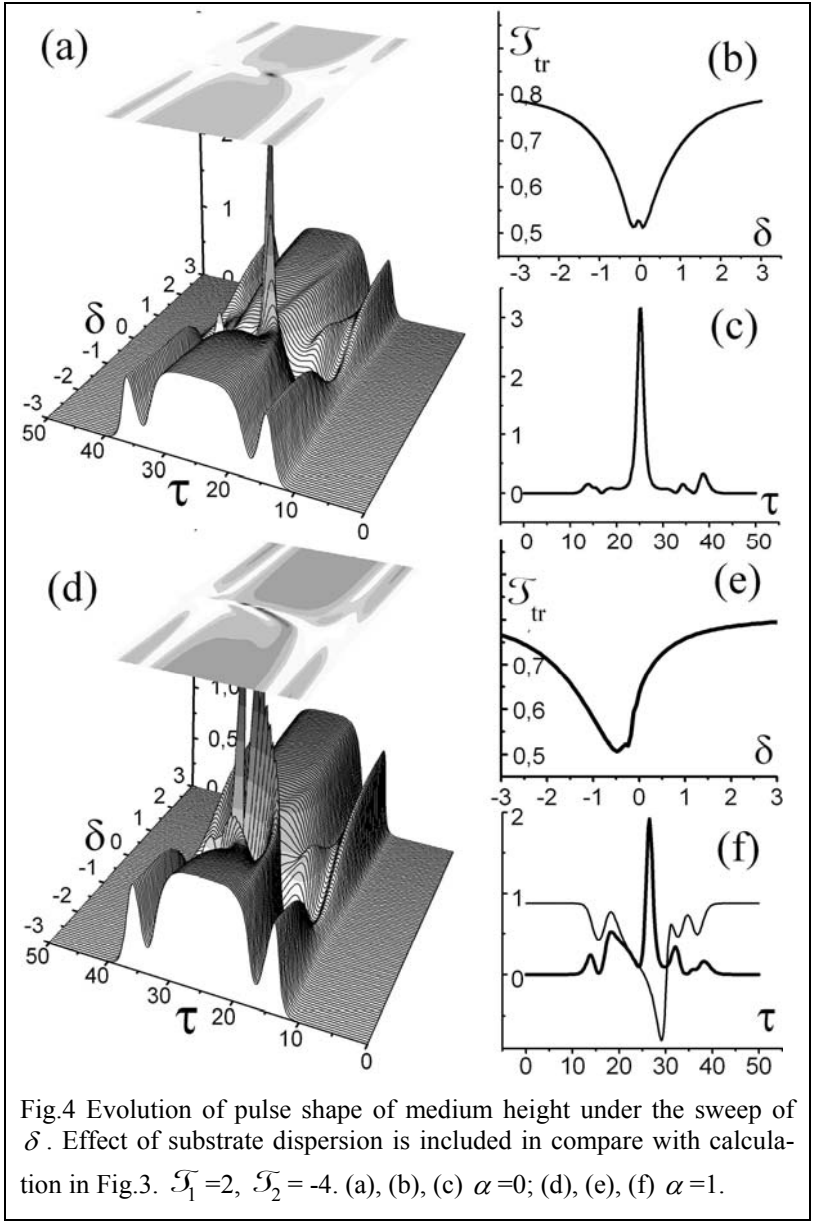

A microscopic Lorentz field introduces an additional dynamic contribution to the total detuning $\delta_{tot}(\tau) = \delta + \alpha n(\tau)$, which is proportional to the resonance level population difference. In consequence, the process of strong coherent interaction of radiation with atomic ensemble starts earlier at negative detuning and the region of coherent interaction becomes wider in $\delta$ domain (Fig.3(d)). The number of spikes also increases (Fig.3(d)) as new peaks appear every time the total detuning changes its sign (these moments are marked by two spots in Fig.3(f.)). The enhancement of a local field effect contribution apparently leads to the distortion of the transmission coefficient

Fig.4 Evolution of pulse shape of medium height under the sweep of $\delta$. Effect of substrate dispersion is included in compare with calculation in Fig.3. $\mathscr{T}_1$ =2, $\mathscr{T}_2$ = -4. (a), (b), (c) $\alpha$ =0; (d), (e), (f) $\alpha$ =1.

$\mathscr{T}_{tr}$ spectrum. There is a slight shift to the negative side and the broadening of the above quoted maximum (Fig. 3(e)).

## D. The effect of substrate dispersion

The effect of substrate material dispersion strongly depends on the value and the sign of the dispersion coefficients $\mathscr{T}_1$ and $\mathscr{T}_2$, leading to a form distortion (Fig. 4(a,b,e,f)) and spectrum changing (Fig.5). The numerical simulation in this paragraph exhibits one certain realization of a dispersion effect from a set of possibilities.

In Fig.4, which is analogus to Fig.3, the effect of substrate dispersion is presented for two cases when the local field effect is off ($\alpha$=0) (Fig. 4(a,b,c)) and when it is on ($\alpha$=1) (Fig. 4(d,e,f)).

The region of resonance, where we drag the quantum system in by sweeping the detuning $\delta$, is getting wider due to the change of spectrum of the refracted field. The coherent interaction of field with the film manifests in a sharp spike of transmitted radiation (Fig. 4(c)) in the vicinity of resonance.

The layer on a dispersive substrate becomes dark in the transmitted light (Fig. 4(b)). With the including of local field effect (Fig. 4(d,e,f)) the population difference provides the dynamical contribution to a total detuning $\delta_{tot}$, making it the time depending value (Fig. 4(f)) and provoking a spectrum distortion (Fig. 5). Consequently the temporal profile of a transmitted field gets more complex in the region of effective coherent interaction (Fig.4 (d,e,f)).

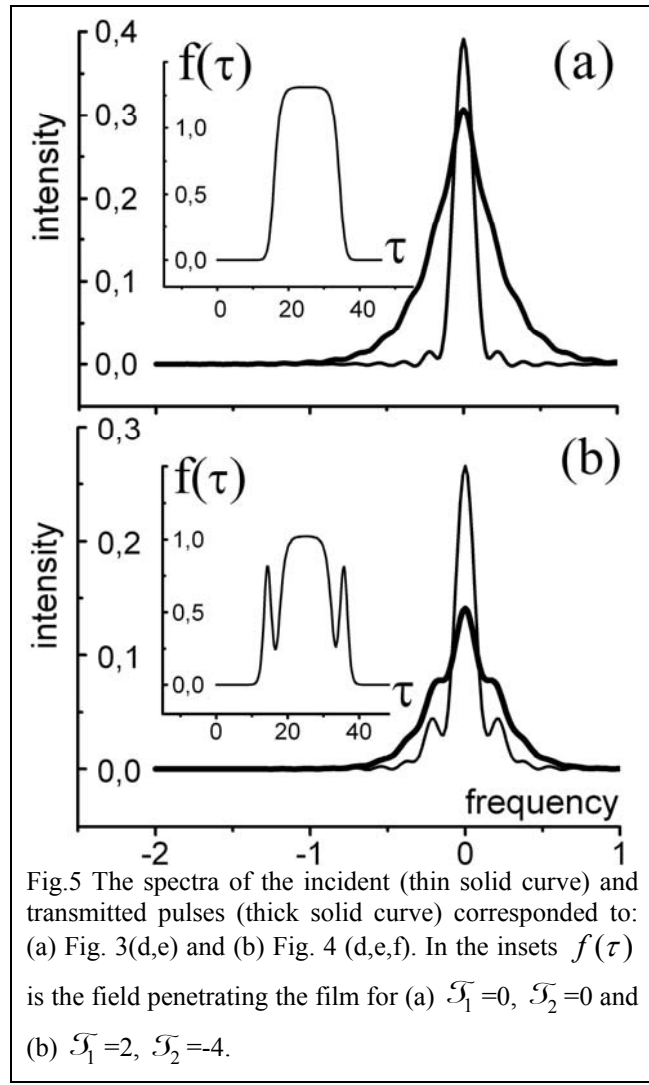

Fig.5 The spectra of the incident (thin solid curve) and transmitted pulses (thick solid curve) corresponded to: (a) Fig. 3(d,e) and (b) Fig. 4 (d,e,f). In the insets $f(\tau)$ is the field penetrating the film for (a) $\mathcal{F}_1$ =0, $\mathcal{F}_2$ =0 and (b) $\mathcal{F}_1$ =2, $\mathcal{F}_2$ =-4.

### 3. PHOTON ECHO IN A THIN FILM OF RESONANCE ATOMS

The photon echo[21] is a coherent response of an ensemble of quantum resonance systems (two- or more level atom) to by a series of ultra-short pulses. Photon echo arises at the moments multiple to time intervals between pulses as the result of phase synchronization of individual radiators with detuning $x$ comprising the inhomogeneous line. The latter process is represented by the averaging $<...>$ in the set of basic equations (11). The difficulty in solving (11) is that the field which acts on atom $e_{loc}(\tau)$ in its turn does depend on averaged polarizability $<\sigma_x>$, that makes the problem self-consistent.

So far there have been several approaches to resolve the problem by admitting the simplifying suggestions[2,6,18]. We report the results of one more attempt to tackle the problem in full including local field effect, inhomogeneous broadening and dispersion of substrate. Equations (11) were solved with the desired accuracy by iterations over field, which started from the field entering the film. The length of the Bloch vector was monitored at every iteration step and at every point of time grid.

In Fig. 6(a) effect of photon echo is depicted with parameter $g$ formally equals to zero. That corresponds to an approximation of given field. The signal of primary echo is delectated in polarization, so the field of a reciprocal reaction of resonance medium is negligible. The picture of multiple echo in Fig. 6(b), where $g{\neq}0$, (echo from echo) is typical for coherent effects in optically dense medium[24].

The local field effect (Fig.6(c)) causes dynamical shifts of spectral components

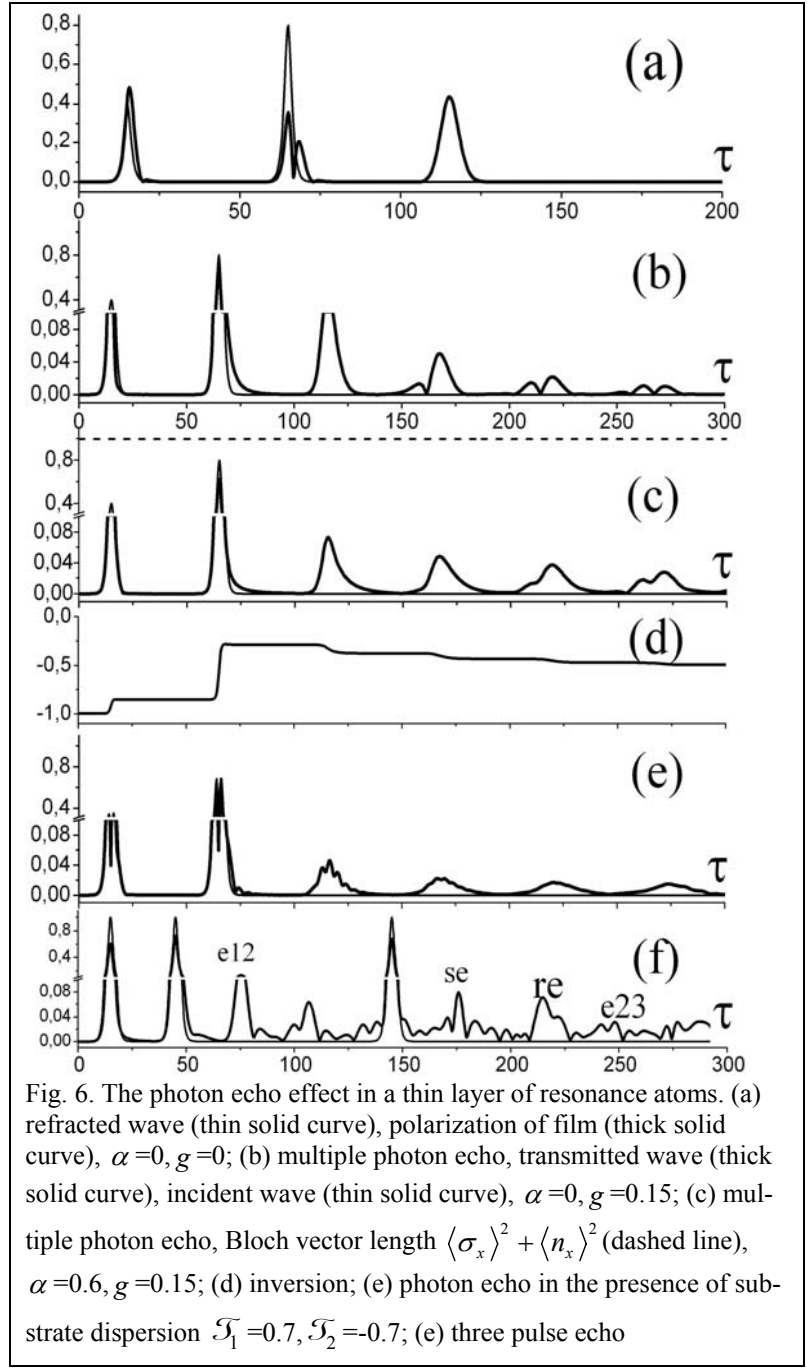

Fig. 6. The photon echo effect in a thin layer of resonance atoms. (a) refracted wave (thin solid curve), polarization of film (thick solid curve), $\alpha$ =0, $g$ =0; (b) multiple photon echo, transmitted wave (thick solid curve), incident wave (thin solid curve), $\alpha$ =0, $g$ =0.15; (c) multiple photon echo, Bloch vector length $\langle\sigma_x\rangle^2 + \langle n_x\rangle^2$ (dashed line), $\alpha$ =0.6, $g$ =0.15; (d) inversion; (e) photon echo in the presence of substrate dispersion $\mathcal{T}_1$ =0.7, $\mathcal{T}_2$ =-0.7; (e) three pulse echo

inside the inhomogeneous line, thus leading to a general smoothing of the time shape of coherent responses. Note that the length of Bloch vector averaged over the spectral line keeps constant (Fig. 6(c) dashed line).

Additional spectral broadening due to substrate dispersion (Fig. 6(e)) acts in the same direction thus depleting the chain of multiple photon echoes. The population difference changes sig-nificantly during the excitation and echo formation, as it is shown in Fig.6 (d). Fig. 6(f) demonstrates the effect of three-pulse excitation attributed by generation of stimulated echo (se) in the appropriate moment[22], as well as reconstructed echo[22] (re), and echoes from the pairs of external pulses. This is a novel result in thin layer optics accounting the local field effect. The substrate dispersion effect for stimulated echo is trivially similar to that in Fig. 6(e).

## CONCLUSION

In conclusion, the numerical analysis of coherent responses of resonance thin film to ultra-short optical pulse excitation is presented. The role of inhomogeneous broadening of resonance line, the local field effect and the substrate dispersion was demonstrated in both the temporal shape of transmitted wave and the integral transmission coefficient. There is a range of incoming amplitudes where the interaction of pulsed light with the quantum system leads to the formation of peaks of superradiation in transmitted wave. Photon echo effect is considered without conventional simplification. It turns out that a film of resonance atoms is able to radiate multiple echo signals after the irradiation by two or more coherent pulses.

## ACKNOWLEDGMENTS

The author gratefully acknowledges fruitful discussions with Andrei I. Maimistov and Askhat M. Basharov. The research is supported by the RFBR grant 06-02-16406.